# Picometre-scale real-time drift correction in TEM and STEM by dynamic control of the specimen stage for atomic-resolution imaging

Christophe Gatel[1*], Julien Dupuy[1], Teresa Hungria[2] and Martin J. Hÿtch[1]

[1] Université de Toulouse, CNRS, CEMES, 31055 Toulouse, France

[2]CNRS, Centre de Microcaracterisation Raimond Castaing, Toulouse, France

*Corresponding author:
christophe.gatel@cemes.fr

## Abstract

In this work, we show how the specimen drift can be actively compensated by controlling the stage with precision down to the picometre scale. We do this by dynamic control, a generic real-time feedback framework designed to actively stabilize electron microscopy experiments by continuously monitoring an experimental variable from the detector data stream and compensating its evolution during acquisition. The framework, applicable to a broad range of controllable experimental instabilities, includes automated calibration procedures, operates in parallel with image acquisition and is implemented as a software plugin without requiring any hardware modification of the microscope.

Results for live drift correction are shown for a selection of TEM and STEM instruments using conventional mechanical stages as well as piezoelectric stages. Experimental results are presented for medium resolution TEM, high-resolution TEM, HR-STEM and *in situ* observations. With piezoelectric stages, specimen stabilization down to the picometre scale was achieved allowing drift-corrected atomic-resolution imaging. Specimen stage-based stabilization significantly improves long-exposure imaging and *in situ* experiments by increasing the effective exposure time, preserving the field of view, maintaining identical optical conditions and eliminating the need for numerical alignment of large datasets. Beyond the specific application presented here, dynamic control provides a versatile framework for real-time regulation of electron microscopy experiments and opens new perspectives for quantitative imaging, automated *in situ* studies and multimodal acquisitions.

## Keywords

TEM, STEM, In situ, Real-time feedback control, Automation, Drift correction, Specimen stabilization

## Introduction

TEM instrumentation has undergone continuous technological development over the past decades. Advances such as aberration correctors[1,2], direct electron detectors[3,4] and improved spectroscopic instrumentation[5,6] have continuously extended the quantitative capabilities of modern electron microscopes by improving their spatial resolution, energy resolution, sensitivity and signal-to-noise ratio (SNR). These developments have not only enhanced conventional imaging and analytical techniques but have also enabled entirely new experimental approaches, including advanced quantitative phase imaging and multidimensional acquisition methods. As a result, an increasing number of TEM experiments now rely on quantitative measurements performed over extended acquisition times and under increasingly complex experimental conditions.

In parallel with these instrumental advances, microscope automation has become an increasingly important area of research. Initially developed for routine operations such as autofocus and astigmatism correction, automation has progressively expanded to higher-order aberration alignment[7], electron tomography[8–12], single-particle cryo-electron microscopy[13], and high-throughput acquisition strategies[14,15]. Similar approaches have also been developed for *in situ* experiments, where image shifts induced by external stimuli are compensated using pre-calibrated microscope responses, allowing predefined experimental sequences to be executed without operator intervention [16]. In such approaches, however, the microscope follows a predetermined acquisition scenario in which the applied corrections are known in advance.

Sequential automation has considerably improved the reproducibility, throughput and efficiency of electron microscopy experiments while reducing operator intervention and electron dose for beam-sensitive materials. Despite these advances, instrumental instabilities remain one of the principal factors limiting the quantitative performance of modern-day electron microscopes. By reducing the achievable signal-to-noise ratio and limiting the effective spatial resolution, they ultimately constrain the full exploitation of recent instrumental developments. In particular, specimen drift remains one of the most pervasive sources of image degradation during long-exposure acquisitions or *in situ* experiments. The current solution, adopted for instance in STEM[17–20] and holographic[21–23] studies, generally consists of recording image series with short exposure times followed by numerical image registration through advanced software specifically developed for this purpose. This strategy is very effective but requires the acquisition and storage of large datasets, increases computational cost, reduces the effective field of view and prone to introducing numerical artefacts. More fundamentally, it raises the question of whether instrumental instabilities could instead be measured continuously and compensated autonomously while the experiment is undergoing.

This concept forms the basis of the dynamic control, a generic real-time feedback framework in which an experimental variable is continuously extracted from the detector data stream and fed back to the microscope to compensate instrumental instabilities during acquisition[24,25]. Although the framework can, in principle, be applied to many measurable experimental variables, including beam position, spectroscopic energy drift or interference fringe position in electron holography, specimen drift constitutes its most universal application because it affects virtually every TEM and STEM experiment. This capability is particularly attractive for *in situ* experiments, where continuously evolving experimental conditions often generate additional instabilities that limit long-duration observations and quantitative measurements.

In a previous work, we demonstrated for the first time the feasibility and the efficiency of real-time feedback control in medium-resolution off-axis electron holography, where simultaneous stabilization of the interference fringes[26] and the specimen position enabled exposure times of several tens of minutes without relying on image stacks[27]. We used a mechanical stage but, more recently, the same feedback principle was implemented on a microscope equipped with a piezoelectric specimen stage, enabling automated specimen stabilization for electron holography and *in situ* observations[28]. However, the demonstrated stabilization remained at the nanometer scale, and its impact on atomic-resolution imaging was not investigated.

Specimen drift can also be corrected in real time using beam-deflector[29–31]. However, beam-deflector compensation modifies the optical conditions during acquisition, whereas mechanical stage stabilization preserves the optical alignment and therefore maintains identical imaging conditions throughout the experiment. To our knowledge, the stabilization performance achievable with piezoelectric specimen stages has not yet been quantitatively characterized for high-resolution imaging, either TEM or STEM.

In the present work, we introduce the dynamic control as a generic real-time feedback framework for (S)TEM and demonstrate its capabilities through active specimen stabilization. Implemented as a software plugin operating in parallel with image acquisition, the framework requires no hardware modification of the microscope and can be readily deployed on modern instruments providing access to detector data streams and microscope control interfaces. Using both conventional mechanical stages and piezoelectric specimen stages on several transmission electron microscopes from different manufacturers, we demonstrate real-time stabilization down to the picometer scale. We further show that active stabilization substantially improves long-exposure TEM and STEM imaging by increasing the effective exposure time, preserving the field of view, suppressing scan distortions and eliminating the need for post-processing image registration. These capabilities are particularly beneficial for *in situ* experiments, where maintaining identical observation conditions over long acquisition times is essential for quantitatively monitoring the evolution of the specimen. Although specimen drift compensation is used here to demonstrate the approach, the same feedback architecture can readily be extended to the active stabilization of many other experimental parameters.

## 1. Dynamical control framework

### a. Feedback loop

The dynamic control is based on feedback units, which provide an efficient means of regulating one or several physical quantities when they can only be controlled indirectly [24,25]. The general architecture of a feedback unit is illustrated in Figure 1a. Although described here in the context of (S)TEM, the same principles are equally applicable to scanning electron microscopy.

The feedback unit receives two inputs: a data stream provided by a camera or a STEM detector, and a setpoint (SP) which corresponds to the reference value of the experimental variable to be maintained. The SP may be defined either by the user or by another process and, depending on the application, may correspond to a specimen position, a beam position or a focus value, for example. A sensor extracts the corresponding process variable (PV) from the incoming data and sends it to the controller. The controller continuously compares the PV with SP and determines the appropriate control output (CO) to be sent to the appropriate microscope component so as to minimize the difference between the two quantities. The component may correspond to a specimen-stage displacement for specimen-position control, a beam-deflector command for beam-position control, or a change in objective-lens excitation for focus control, for example. The control output is applied through a communication library provided by the microscope manufacturer, supplier or developed by the user.

At the end of each iteration, the feedback unit generates a synchronization signal (sync) to request a new frame and initiate the next iteration. Although operating continuously in time, the feedback unit operation is therefore intrinsically discrete and iterative: a physical quantity is measured, compensated for, measured again, and so on. This naturally defines an execution period and an associated operating frequency for the feedback loop.

In the present implementation, both the sensor and the controller are software-based modules. The sensor relies on image processing algorithms to extract the process variable from the detector data stream. The design of the controller is a broader problem, as it requires a model describing the response of the controlled system. Several levels of complexity can be considered depending on the targeted application.

The most widely used example is the proportional-integral-derivative (PID) controller, originally introduced by Minorsky (1922) [32], which can be written as:

$$CO(t) = G_P e(t) + G_I \int_0^t e(t)\mathrm{d}t + G_D \frac{\mathrm{d}e(t)}{\mathrm{d}t} \quad (1)$$

where the error defined as $e$ = PV − SP, and $G_P$, $G_I$ and $G_D$ are the proportional, integral and derivative gains, respectively This controller introduces a response proportional to the measured error while accounting for both its temporal evolution and its history. The derivative term anticipates future deviations from the setpoint, whereas the integral term minimizes residual errors and can accelerate convergence toward the desired value. When appropriately tuned, a PID controller provides rapid convergence without overshoot or oscillations around the setpoint. It is particularly efficient in a wide range of applications, provided that the controller gains are properly calibrated, which is the main drawback of this approach

A simpler alternative consists of using a proportional controller, corresponding to a PID controller with $G_I = G_D = 0$. Its main advantage lies in its simplicity, as it does not require any prior knowledge of the physical state of the controlled system. Proportional control is particularly well suited for compensating small perturbations exhibiting locally linear behaviour around the setpoint, such as instrumental instabilities that do not involve significant inertia or mechanical backlash.

The purely proportional control is used for all the experiments reported in this work. This approach is particularly well suited for compensating small perturbations exhibiting locally linear behaviour around the setpoint, such as instrumental instabilities that do not involve significant inertia or mechanical backlash, provided that measurement and compensation are sufficiently fast to maintain only small deviations between successive iterations of the feedback loop.

### b. Calibration loop

For feedback operation, the relationship between the control output applied to the microscope and the resulting variation of the process variable must first be determined. This is the purpose of the calibration unit shown in Figure 1b. The SP, PV and CO have the same physical meaning as in the feedback unit, and the same sensor is used to measure the process variable. Their role during calibration, however, is different.

Instead of compensating an uncontrolled deviation from the setpoint, the calibration module deliberately modifies the controlled microscope component through the CO and measures the resulting variation of the PV from the detector data stream using the sensor. By comparing the applied CO with the measured variation of the PV, the calibration module determines the numerical relationship required by the controller during subsequent feedback operation, and define the corresponding setpoint as a calibration value that is returned once the calibration procedure has been completed.

For example, when calibrating specimen-position control, a known CO is sent to the specimen stage to produce a displacement along one of its axes. The resulting specimen displacement is measured from the detector frames by the same sensor subsequently used for drift compensation. Repeating this procedure provides the calibration coefficients relating stage commands to specimen displacements in the image.

Once the calibration procedure has been completed, the calibration parameters are stored and transferred to the controller of the feedback unit. The calibration unit is then no longer active during feedback operation.

The calibration procedure is a key component of the dynamic control framework and must be considered as an integral part of the development of any feedback loop. It should be sufficiently fast to minimize the experimental overhead while remaining robust over the duration of an experiment in order to reduce the need for repeated calibrations. Particular attention must therefore be paid to both the stability of the calibration parameters and the efficiency of the calibration algorithms.

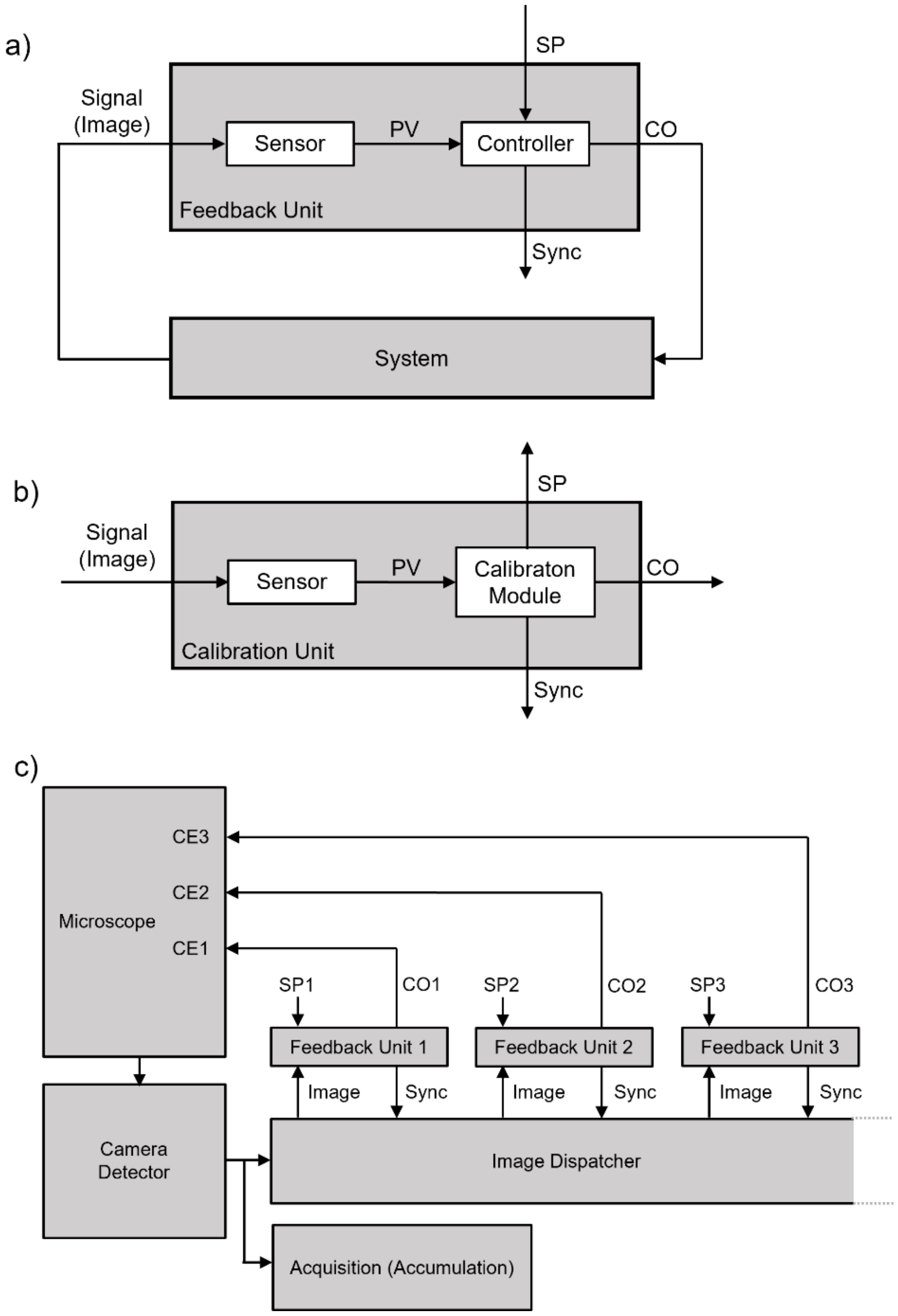

***Figure 1.*** *(a) Architecture of a feedback unit. The sensor extracts the process variable (PV) from the detector data stream and transmits it to the controller, which continuously compares it with the setpoint (SP) and generates the corresponding control output (CO). A synchronization signal (sync) is produced at the end of each iteration to initiate the next loop cycle. (b) Architecture of the calibration unit used to determine the calibration coefficients required by the controller. SP and CO identical to the feedback unit, but are deliberately varied by the calibration module to*

*characterize the response of the controlled system. The resulting calibration parameters are subsequently used by the controller during feedback operation. (c) Dynamic control implementation in which several feedback loops operate simultaneously during an experiment. Calibration Units are not represented because calibration is performed beforehand and calibration units are inactive during feedback operation.*

### c. Auto-oscillations and parallel compensation

A feedback loop may generate self-oscillations that manifest themselves as an instability of the controlled system. In such cases, the process variable (PV) exhibits increasingly large oscillations around the setpoint. For proportional control, this behaviour typically arises from two situations. The first occurs when the calibration value is underestimated with respect to its actual value. As a consequence, the controller generates an overestimated control output at each iteration, resulting in a larger deviation from the setpoint after compensation. The feedback loop then overcompensates the error at the following iteration, ultimately behaving as an amplified oscillator. The second situation originates from an inappropriate definition of the loop execution period $t_{loop}$. The next iteration may start before the previous compensation has been fully applied, either because the controlled element has not yet reached its new state or because the loop frequency exceeds the time required to acquire a new frame. In this case, the feedback loop measures a process variable that does not yet include the effect of the previous compensation and therefore applies an additional correction of similar magnitude. This overcompensation drives the system away from the setpoint until a subsequent iteration brings it back close to equilibrium, resulting in an oscillatory behaviour.

Particular attention must therefore be paid both to the accuracy of the calibration procedure and to the synchronization between the feedback loop and the detector data stream. The frame analysed at a given iteration must always contain the compensation applied during the previous iteration. Consequently, the loop execution period $t_{loop}$ must respect the following condition:

$$t_{loop} \geq t_{frame} + t_{compute} + t_{response} \tag{2}$$

with $t_{frame}$ the frame acquisition time, $t_{compute}$ the computation time (sensor, controller, …) and $t_{response}$ the physical response time of the controlled element. Optimizing the image processing algorithms and using fast and sensitive detectors therefore directly improves the responsiveness of the dynamic control by allowing higher loop frequencies.

A consequence of the Eq. (2) is that no universal maximum correctable drift rate can be defined. The practical limit depends on the drift accumulated during both a single frame and a complete feedback-loop iteration relative to the spatial accuracy required for the experiment. It therefore depends on the detector frame rate and sensitivity, the size of the image region processed by the sensor, the response and positioning precision of the controlled element, and the targeted spatial resolution.

Several feedback units can operate simultaneously, either independently or synchronously, each regulating a different process variable with their own $t_{loop}$ (Figure 1c). This approach has been implemented for electron holography experiments, where specimen drift and interference fringe drift are compensated independently by two dedicated feedback units while a third unit performs image acquisition[27]. In such configurations, an appropriate resource-sharing strategy is required to ensure non-concurrent access to both the microscope controls and the detector data stream. In Figure 1c, this task is performed by the image dispatcher, which distributes the detector frames to the different feedback units.

The dynamic control can be implemented on any electron microscope capable of providing a detector data stream and allowing access to at least one controllable experimental parameter. Its performance naturally benefits from high detector frame rates and improved sensitivity. The latest generation of direct electron detectors, capable of delivering tens to hundreds of high-resolution frames per second, as well as fast STEM scans, therefore provide particularly favourable conditions for the implementation of efficient and highly responsive live compensations of instabilities.

In the present work, only the feedback loop dedicated to specimen stabilization is described in detail. However, many of the concepts presented here are directly applicable to other dynamic control implementations targeting different experimental instabilities.

## 2. Feedback loop for real-time sample drift compensation

Real-time compensation of specimen drift can be implemented by acting directly on the microscope using two different approaches:

- mechanically, by moving the specimen using the specimen stage;
- electronically, by translating the image using beam deflectors (deflector-shift-based compensation).

The specimen stage provides the most straightforward and natural means of controlling the specimen position. It allows large displacements while preserving the optical conditions of the microscope. However, when based on a purely mechanical design, the precision in displacement and sensitivity are limited. Mechanical stages also exhibit inertia and therefore respond more slowly than electron-optical components. Large excitations may also induce non-linear behaviour and mechanical backlash, which are difficult to control accurately. In principle, atomic-resolution imaging is therefore only possible with piezo-stage control.

Image translation can be achieved using beam deflectors located in an image plane below the objective lens for TEM[29] or in the scan system above the sample in STEM[30,31]. These deflectors provide a fast and precise response. Their main limitation is that they modify the electron-optical conditions by translating the entire electron beam. Specimen drift is therefore converted into a beam displacement, requiring careful alignment of the microscope optics to avoid unwanted beam tilts with respect to the optical axis. They also may exhibit a behaviour that deviates from linearity for large shift values.

Both approaches present complementary advantages and limitations. The developments presented here are common to both stage-based and deflector-shift-based compensation schemes. However, all the experimental results reported in this work were obtained using stage-based compensation only, firstly with a mechanical stage for medium-resolution imaging and then and piezoelectric stage for atomic-resolution imaging.

### a. Sensor

Specimen drift is measured by comparing the latest frame provided by the camera or STEM detector with a reference frame. In the following, $x$ and $y$ denote the horizontal and vertical image axes, respectively, whereas $X$ et $Y$ correspond to the specimen stage axes. The image features used for this comparison must of course remain unchanged between the two frames. The reference frame may correspond to the first frame acquired at the beginning of the experiment, be updated by the user during the experiment, or simply correspond to the frame acquired during the previous iteration of the feedback loop.

The drift measurement is performed using cross-correlation. The position of the correlation peak with respect to the center of the cross-correlation image provides the translation vector $\begin{pmatrix} \Delta x \\ \Delta y \end{pmatrix}$, whose coordinates correspond to the specimen drift along the $x$ and $y$ image directions. Starting from the integer-pixel position of the correlation peak, sub-pixel precision can be achieved using standard approaches such as centre-of-mass determination or polynomial and Gaussian fitting methods.

The main challenge faced by the sensor is the identification of image features that allow accurate and robust drift measurements. Cross-correlation is highly sensitive to image contrast and is therefore predominantly influenced by the most contrasted features present in the frame. The user may select a reduced region of interest exhibiting sufficient contrast in order to improve both the measurement precision and the computational speed. However, when very short frame times are used, image contrast can become dominated by high-frequency Poisson noise arising from the limited number of collected electrons. Under such conditions, the cross-correlation result may become unreliable or even completely random. To overcome this limitation, two main types of digital filters can be applied before cross-correlation. Low-pass filters, such as Butterworth or Gaussian filters, are used to suppress high-frequency noise, whereas derivative filters, such as the Sobel filter, enhance the contours of relevant image features. These filtering procedures improve the robustness of the cross-correlation measurement, particularly at low electron doses and short frame times. As no universal filter exists, it is important that a selection of filters be readily available so that they can be tested during the microscope session. The most efficient filter for the experiment at hand can then be designated for the calibration and feedback control.

**b. Controller**

As discussed above, specimen drift compensation and the associated calibration procedure are implemented by acting on the specimen stage, whereas the drift measurement is performed on the detector frames.

The specimen displacement measured in the image $\begin{pmatrix} \Delta x \\ \Delta y \end{pmatrix}_{\text{image}}$ is related to the actual specimen displacement $\begin{pmatrix} \Delta x \\ \Delta y \end{pmatrix}_{\text{objet}}$ through the magnification relationship:

$$\begin{pmatrix} \Delta x \\ \Delta y \end{pmatrix}_{\text{image}} = M\boldsymbol{R}_M \begin{pmatrix} \Delta x \\ \Delta y \end{pmatrix}_{\text{objet}} \tag{3}$$

where $M$ is the linear magnification and $\boldsymbol{R}_M$ is the rotation matrix arising, for example, from the Larmor rotation induced by the magnetic lenses located between the specimen and the detector.

Similarly, the relationship between the specimen displacement and the displacement applied by the specimen stage $\begin{pmatrix} \Delta X \\ \Delta Y \end{pmatrix}_{\text{stage}}$ can be written as:

$$\begin{pmatrix} \Delta x \\ \Delta y \end{pmatrix}_{\text{objet}} = \boldsymbol{P}_{\text{stage}} \begin{pmatrix} \Delta X \\ \Delta Y \end{pmatrix}_{\text{stage}} \tag{4}$$

where $\boldsymbol{P}_{\text{stage}}$ is the basis transformation matrix between the specimen stage reference frame, which is neither necessarily orthogonal nor normalized, and the orthonormal reference frame attached to the specimen.

Rather than relying on an analytical expression for these quantities, Eqs. (2) and (3) can be combined into the simpler relationship:

$$\begin{pmatrix}\Delta x\\ \Delta y\end{pmatrix}_{\text{image}} = M\boldsymbol{R}_M\boldsymbol{P}_{\text{stage}}\begin{pmatrix}\Delta X\\ \Delta Y\end{pmatrix}_{\text{stage}} = \boldsymbol{C}_{\text{stage}}\begin{pmatrix}\Delta X\\ \Delta Y\end{pmatrix}_{\text{stage}} \quad (5)$$

The matrix $\boldsymbol{C}_{\text{stage}}$ corresponds to the calibration matrix used by the feedback unit.

At each iteration of the feedback loop, the controller must apply the displacement required to compensate the measured specimen drift, i.e. $\begin{pmatrix}-\Delta x\\ -\Delta y\end{pmatrix}_{\text{image}}$, to maintain the specimen at its reference position. The control output sent to the specimen stage is therefore given by:

$$\begin{pmatrix}\Delta X\\ \Delta Y\end{pmatrix}_{\text{stage}} = \boldsymbol{C}_{\text{stage}}{}^{-1}\begin{pmatrix}-\Delta x\\ -\Delta y\end{pmatrix}_{\text{image}} \quad (6)$$

The specimen stage is digitally controlled and consequently operates with discretized displacement values. The stage motion is defined along its two axes $X$ and $Y$ by minimum displacement steps $s_X$ and $s_Y$. The actual compensation $\begin{pmatrix}n_X s_X\\ n_Y s_Y\end{pmatrix}_{\text{stage}}$ applied at time $t$ therefore corresponds to the closest integer number $n_X$ and $n_Y$ of minimum displacement steps that can be generated by the stage controller. The resulting compensation error $\sigma_{stage}$ is consequently bounded by half of the minimum displacement step along each stage axis.

Although the measured specimen drift scales with magnification, it generally remains sufficiently small when evaluated over short time intervals for the required compensation to be limited to one or a few minimum displacement steps, or even no compensation at all. The dynamic control therefore consists in determining the appropriate instant at which these minimum displacement steps must be applied to maintain the specimen at its reference position.

For mechanical specimen stages, these small and regularly applied displacements naturally favour a linear response of the stage. Moreover, by continuously compensating the specimen drift, the feedback loop effectively counteracts the physical mechanisms responsible for the drift and progressively stabilizes the specimen over time. Consequently, the number of compensations required generally decreases as the experiment proceeds.

Larger specimen drifts may occur during *in situ* experiments involving external stimuli such as temperature variations or mechanical strain. In such situations, the frame time $t_{frame}$ must be reduced in order to minimize the specimen displacement occurring between successive frames and preserve image sharpness. This requirement naturally comes at the expense of the number of collected electrons per frame and therefore of the contrast to be detected by the sensor. The use of fast and highly sensitive direct electron detectors is particularly advantageous under these conditions.

### c. Calibration module

The $X$ and $Y$ specimen stage axes are calibrated independently using the same automated procedure. Let us consider, for instance, the calibration of the $X$ stage axis. The user first defines the number of measurements $N_X$ to be performed and selects an integer multiple $n_X$ of the minimum digital step $s_{digitX}$, chosen to produce a displacement sufficiently large to be accurately measured by the sensor. A reference frame is initially acquired before the calibration unit applies a digital command corresponding to the selected stage displacement. After the delay $t_{frame}$, a second frame is acquired and the induced displacement $\Delta x$ and $\Delta y$ is measured by the sensor along the $x$ and $y$ image directions, respectively. This procedure is repeated $N_X$ times, each newly acquired frame becoming the reference frame for the subsequent iteration.

However, the measurements may contain a contribution from the intrinsic specimen drift occurring during the calibration procedure. To remove this contribution, a second series of N_X measurements is performed using stage displacements of the same magnitude along the same axis but with the opposite sign. Assuming that the parasitic specimen drift remains approximately constant over the calibration time, its contribution has the same sign in both measurement series, whereas the displacement generated by the calibration unit changes sign. Combining the two series therefore cancels the drift contribution and isolates the displacement produced by the stage command, resulting in a round-trip calibration sequence.

The calibration coefficients $c_{x,X}$ and $c_{y,X}$ associated with the $X$ stage axis are subsequently obtained from the mean values of the two measurement series:

$$c_{x,X} = \frac{1}{2N_X}\sum_{i=1}^{2N_X} \frac{\Delta x_i}{n_X s_{digitX}} \quad \text{and } c_{y,X} = \frac{1}{2N_X}\sum_{i=1}^{2N_X} \frac{\Delta y_i}{n_X s_{digitX}} \tag{8}$$

The uncertainty associated with these coefficients corresponds to the statistical uncertainty obtained from the measurements repeated $2N_X$ times.

The same procedure is then applied to the $Y$ stage axis, yielding the corresponding calibration coefficients $c_{x,Y}$ and $c_{x,Y}$. The calibration matrix used by the controller can therefore be written as:

$$\boldsymbol{C}_{\text{stage}} = \begin{pmatrix} c_{x,X} & c_{x,Y} \\ c_{y,X} & c_{y,Y} \end{pmatrix} \tag{9}$$

This matrix directly provides the displacement, expressed in image pixels along the $x$ and $y$ directions, produced by a single minimum displacement step applied to the $Y$ and $X$ stage axes.

The total calibration time typically ranges from one to three minutes depending on the selected frame time and the number of measurements defined by the user. The calibration coefficients depend not only on the microscope magnification but also on the detector sampling conditions, for example when image binning is used. Performing a calibration before each experiment is not necessarily required. Calibration matrices obtained during previous experiments may be recorded and reused if the experimental conditions remain unchanged.

## 3. Dynamic control implementation on a HF3300 Hitachi microscope

The dynamic control framework described in this work was initially developed on the I2TEM microscope, a Hitachi HF3300-C transmission electron microscope specifically designed for *in situ* and interferometry experiments [27]. The microscope is equipped with a cold field-emission gun (CFEG) providing optimal brightness and features a dual-stage configuration consisting of an upper (Lorentz) specimen stage positioned above the objective lens for field-free observations, and a conventional specimen stage located between the objective lens pole pieces. Both stages present the same characteristics. The microscope is equipped with a CEOS BCOR aberration corrector enabling both on- and off-axis aberrations to be corrected in either Lorentz or conventional imaging modes. Spatial resolutions of 0.5 nm and 87 pm can be achieved in Lorentz and conventional high-resolution TEM modes, respectively [33].

A OneView camera (Gatan Inc.) was initially used. The detector consists of a 4k CMOS sensor (4096 × 4096 pixels) with a pixel size of 15 µm and provides up to 25 frames per second at full resolution, corresponding to a minimum frame time of 40 ms and a data throughput of approximately 1.5 $GB.s^{-1}$. In practice, frame times ranging from 0.2 s to 1 s were typically used. More recently, the OneView camera has been replaced by a K3 direct electron detector (Gatan Inc.) featuring a 5760 × 4092 pixel sensor with a 5 µm pixel size. Owing to its higher sensitivity, the K3 detector allows significantly shorter frame times to be used, thereby improving the responsiveness of the dynamic control framework if necessary.

Both cameras are controlled through DigitalMicrograph (Gatan Inc.). The microscope components are interfaced directly within DigitalMicrograph using a dedicated communication library developed in-house, although the native DigitalMicrograph functions can also be used. The graphical user interface, image processing algorithms and dynamic control routines have been entirely implemented within DigitalMicrograph, providing direct access to the detector data stream. Consequently, the dynamic control framework can be readily installed and used within DigitalMicrograph 3.xx.

### a. Specimen stage calibration

The first step in implementing the dynamic control on the I2TEM microscope consisted in calibrating the specimen stage. Figure 2a presents the results obtained from a calibration procedure performed using 2 × 30 measurements (round-trip sequence) for each stage axis and a displacement corresponding to twice the minimum digital displacement step. Images were acquired using the upper specimen stage in Lorentz mode with a frame time ($t_{frame}$) of 0.5 s. The total calibration time for both stage axes was approximately 90 s. The image displacements $\Delta x$ and $\Delta y$ measured along the $x$ and $y$ directions resulting from stage displacements along the $X$ axis are represented in blue, whereas those corresponding to the $Y$ axis are shown in orange.

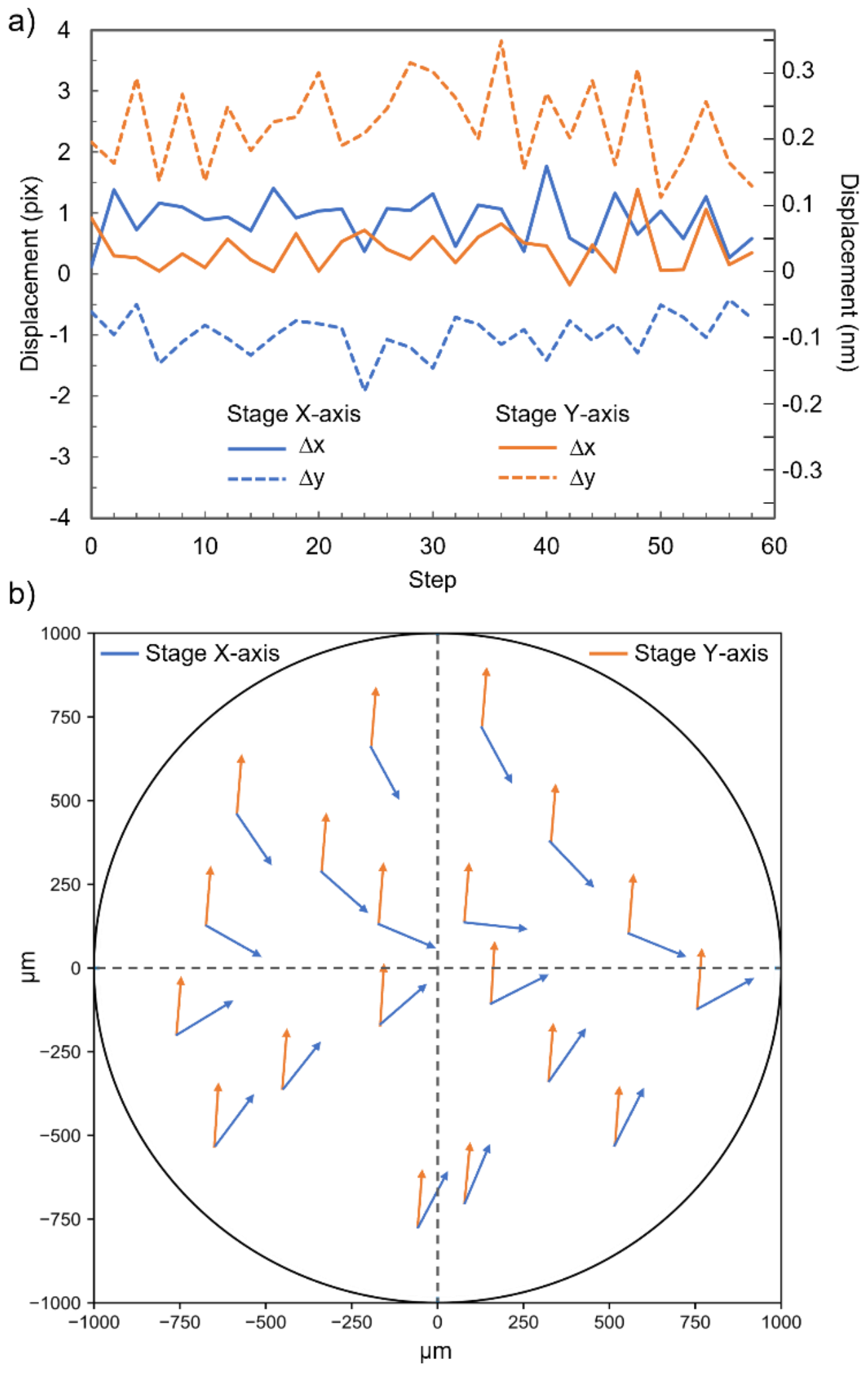

***Figure 2.*** *(a) Calibration curves obtained on a specimen stage of the I2TEM microscope (pure mechanical design). (b) Evolution of the orientation of the specimen stage axes as a function of the stage position, revealing the non-orthogonality and position dependence of the mechanical stage geometry.*

The mean values of the repeated measurements provide the displacement vectors in the image associated with the minimum displacement steps of the specimen stage. For the $X$ axis, the minimum displacement $s_X$ corresponds to 1.32±0.01 pixels and forms an angle of −48±3° with respect to the horizontal image axis. For the $Y$ axis, the measured displacement $s_Y$ is 2.45±0.01 pixels with an angle of 89±2°. The minimum displacement values expressed in pixels naturally depend on the microscope magnification and must therefore be measured or recalculated for each magnification. In contrast, the displacement values at the specimen level are independent of both the magnification and the imaging mode. Knowing the image calibration at the selected magnification (here 0.0937 nm.pixel$^{-1}$), we determined the corresponding minimum displacement steps at the specimen level as $s_X$ = 0.12±0.01 nm et $s_Y$ = 0.24±0.01 nm. Two particularly noteworthy observations arise from these calibration measurements.

First, the minimum displacement steps are remarkably small for a purely mechanical specimen stage, enabling sub-nanometre control of the specimen position. These values are more than one order of magnitude smaller than those specified by the manufacturer. This behaviour may be attributed to the calibration procedure itself. The stage is driven entirely through software commands without any manual intervention, and only the smallest possible digital displacement values are applied. Consequently, the stage is operated in a regime of very small successive linear displacements, which significantly improves its positioning precision.

Secondly, the two stage axes are clearly not orthogonal. The measured angle between them is approximately 40°. Additional calibration measurements performed at different stage positions across the full displacement range (±1 mm) revealed that this angle varies with the stage position. The orientations of the $X$ and $Y$ axes are shown in Figure 2b. The orientation of the $X$ axis exhibits a strong dependence on the $Y$ stage position, the angle between both axes decreasing to approximately 20° when the stage is displaced by 750 µm along the $Y$ direction. No significant variation was observed as a function of the $X$ stage position. This behaviour originates from the mechanical design of the specimen stage.

Similar measurements performed on the conventional specimen stage of the I2TEM microscope revealed comparable characteristics. Beyond providing the calibration parameters, the calibration procedure therefore offers valuable information regarding the actual displacement behaviour of the specimen stage and provides a detailed characterization of its mechanical properties.

### b. Sample drift correction at medium resolution

Figure 3 presents the results obtained with and without dynamic control for long-exposure observations performed under identical imaging conditions to those used for the calibration procedure. The specimen consists of a Ni nanowire with a diameter of approximately 150 nm[27]. The 4096 × 4096 pixel frames were acquired using the OneView camera with a frame time of 0.25 s. A region of interest containing the nanowire tip was selected by the sensor to measure the specimen drift at a frequency of approximately 3 to 4 measurements per second.

During a total acquisition time of 300 s, the 1200 frames provided by the camera data stream were summed in real time to produce the final images shown in Figures 3a and 3b, corresponding to the cases without and with specimen drift compensation, respectively. In the absence of specimen drift compensation, the nanowire edges appear significantly blurred, revealing the degradation of the effective spatial resolution during the acquisition. In contrast, when the feedback unit for drift compensation is activated, the nanowire edges remain sharp and well defined. The corresponding intensity profiles extracted across the nanowire edge are presented in Figure 3c. Without compensation, the specimen drift broadens the edge profile over a distance of approximately 14 nm (100 pixels). When the feedback unit is activated, the edge transition becomes significantly sharper, exhibiting a width of approximately 7 nm (50 pixels), and the native surface oxide layer becomes clearly visible.

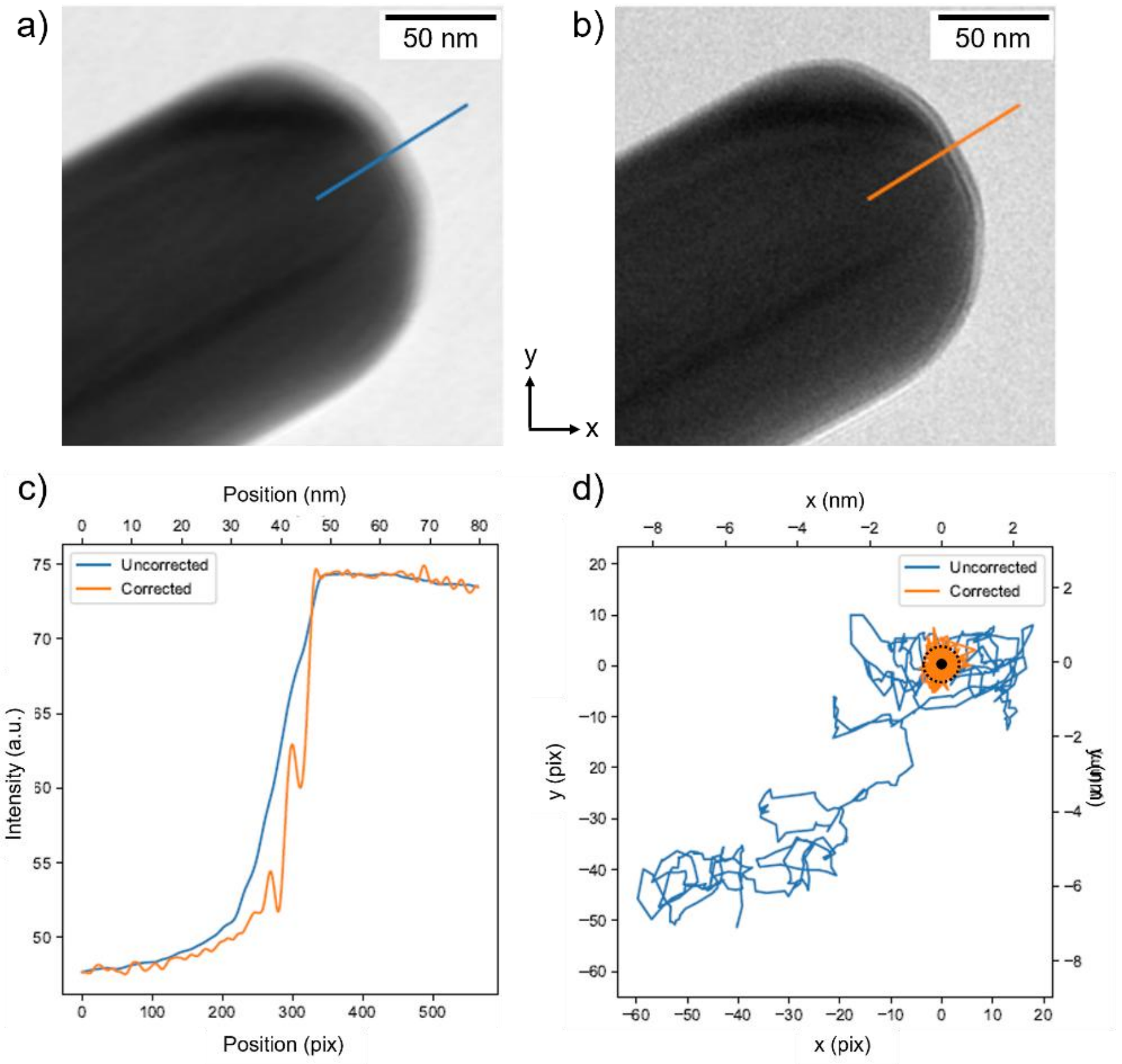

***Figure 3.*** *Long-exposure image (300 s total acquisition time) of a Ni nanowire acquired (a) without specimen drift compensation and (b) with specimen drift compensation. (c) Intensity profiles extracted across the nanowire edge in (a) and (b) along the blue and orange lines, respectively. (d) Specimen drift trajectories measured during the acquisitions.*

The sensor remained active throughout the experiment in both configurations, allowing the specimen drift to be recorded over the entire acquisition time. The corresponding drift trajectories are shown in Figure 3d. The black dot indicates the initial specimen position at the beginning of the experiment, whereas the radius of dashed circle corresponds to the spatial resolution achieved in Lorentz mode under the present experimental conditions (0.5 nm). Without compensation, the specimen drift exhibits a random behaviour and progressively moves the specimen away from its initial position. After 300 s, the total displacement reaches approximately 8 nm. Although relatively small over several minutes, this drift is sufficient to induce the loss of sharpness observed in Figure 3a, even at medium resolution. When the feedback loop is activated, the specimen remains localized around its initial position, with approximately 90 % of the measured positions lying within the dashed circle.

The residual positioning uncertainty obtained using stage-based compensation is therefore smaller than the spatial resolution of the microscope in Lorentz mode. This result may appear counterintuitive at first sight, but it simply reflects the fact that spatial resolution and displacement sensitivity are fundamentally different quantities. The specimen displacement is measured by cross-correlation, which is sensitive to image translations and does not require the individual image features to be spatially resolved. Its ultimate precision is therefore determined by the detector sampling and the sub-pixel accuracy of the sensor algorithms. The overall compensation accuracy is determined by the combined contributions of the sensor measurement uncertainty, the calibration uncertainty of the feedback loop and the finite minimum displacement steps of the specimen stage (see above).

The present implementation only compensates specimen drift within the image plane. Drift occurring along the electron beam direction results in a variation of the focal conditions. The accurate measurement of the focus variation is more challenging but the dynamic control approach is perfectly adapted for stabilizing the desired focal conditions throughout the experiment. Fine real-time focus adjustments can also be performed manually by modifying the current of the objective or Lorentz lens.

Stage-based specimen stabilization implemented on the I2TEM microscope is therefore particularly well suited for medium-resolution observations performed in Lorentz mode. In principle, the specimen position can be maintained indefinitely, allowing virtually unlimited total exposure times provided that the specimen itself does not evolve under electron irradiation or contamination.

Depending on the experimental conditions and the detector frame rate, the feedback unit for drift compensation can operate at frequencies approaching 10 iterations per second. In practice, however, one or two compensations per second are generally sufficient to maintain the specimen position owing to the relatively rather slow dynamics of specimen drift.

Since its implementation on the I2TEM microscope, the dynamic control has been successfully employed in numerous studies[34–41], both to improve the signal-to-noise ratio through long-exposure acquisitions and to maintain a constant field of view during extended *in situ* experiments without user intervention. For example, the specimen position has been actively stabilized for more than four hours during *operando* biasing experiments performed by electron holography.

When the minimum displacement steps of the specimen stage become larger than the spatial resolution of the imaging mode, as is the case in high-resolution TEM where the I2TEM microscope reaches a spatial resolution of 0.08 nm, two alternative approaches can be considered:

- Deflector-shift-based compensation offers highly precise, linear and stable image translations, albeit at the expense of small variations in the electron-optical alignment conditions. These variations remain negligible for small specimen drifts but may become significant during *in situ* experiments involving large temperature-induced displacements. The achievable precision is nevertheless excellent, with minimum image displacements as small as 5 pm measured on the I2TEM microscope.
- Piezoelectric specimen stages or dedicated piezoelectric specimen holders provide significantly finer and more accurate positioning capabilities while preserving the illumination conditions of the microscope. This approach constitutes the basis of the picometer-scale specimen stabilization presented in the following section.

## 4. Picometre-scale specimen stabilization

Several TEM-STEM microscopes are equipped with piezoelectric specimen stages offering positioning precisions that are well suited for picometer-scale stabilization. Among them, the JEOL JEM-ARM and JEM-NeoARM microscope families provide particularly attractive platforms for real-time specimen stabilization. The dynamic control framework described in this work has been successfully implemented and tested on three such microscopes (two JEM-ARM200F microscopes and one JEM-NeoARM200F microscope), each providing direct access to their cameras and STEM detectors through DigitalMicrograph. The communication libraries required to interface the microscope controls were natively available within DigitalMicrograph.

The calibration procedure revealed nearly identical stage characteristics for the three microscopes. Several calibration matrices were therefore established to enable the feedback unit operation in both TEM and STEM imaging modes using cameras and High-Angle Annular Dark Field (HAADF) or Annular Bright Field (ABF) detectors.

Figure 4a presents a representative calibration on one of the microscopes. In contrast to the mechanical specimen stage of the I2TEM microscope, the calibration curves are considerably more stable and exhibit significantly smaller displacement amplitudes. The minimum displacement steps $s_X$ and $s_Y$ of the piezoelectric specimen stages were measured to be 15±1 pm and 5±1 pm, respectively, for the three microscopes investigated, values that are comparable to those achieved using deflector-shift-based compensation. The angle between both stage axes was measured to be 90±3°, demonstrating the excellent behaviour of the piezoelectric stages allowing atomic-scale compensation. We performed measurements at several positions on the piezoelectric stages and confirmed that, unlike for the I2TEM microscope stage, the calibration is independent of stage position.

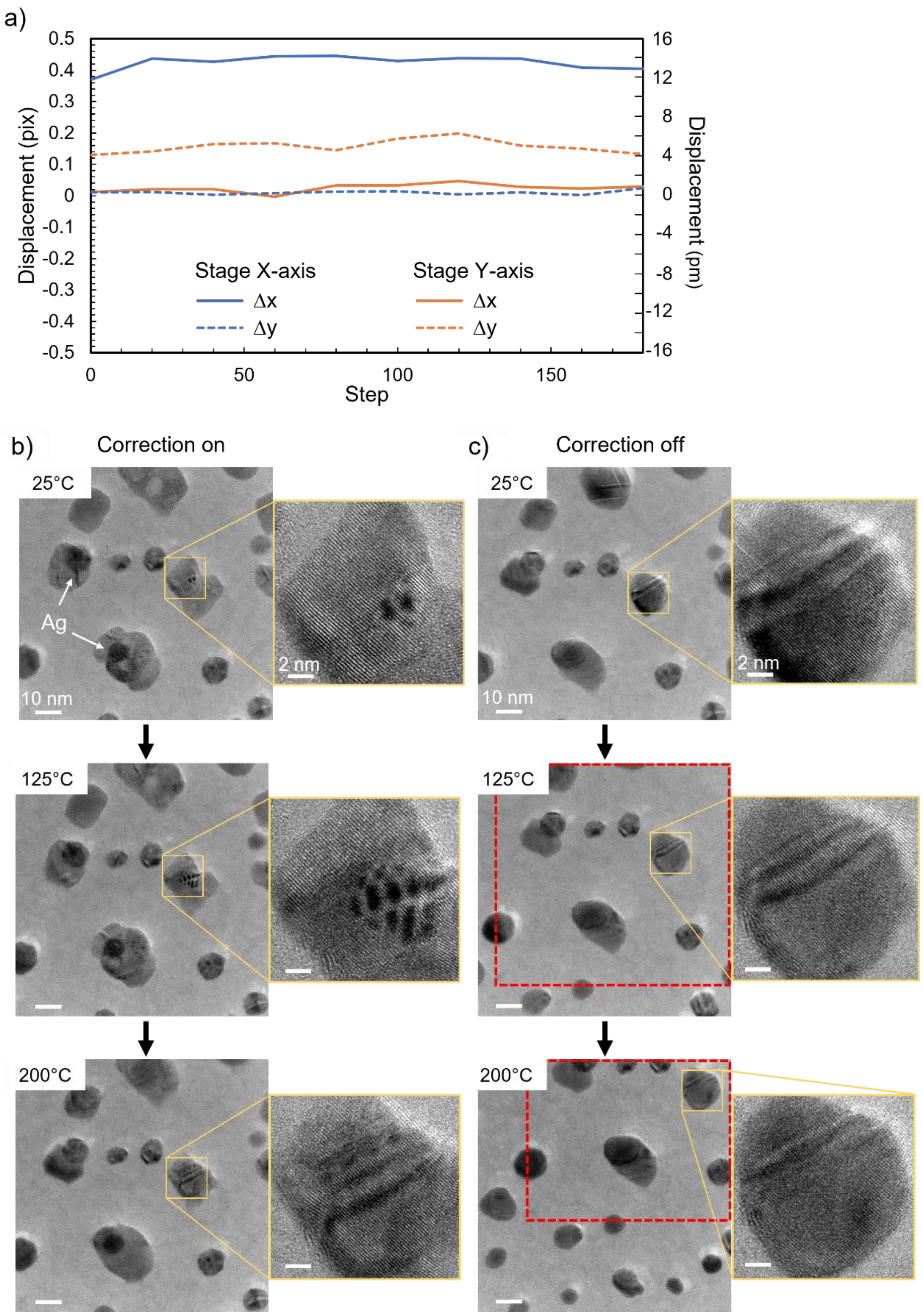

***Figure 4.*** *(a) Calibration curves obtained on a piezoelectric specimen stage of a JEM-ARM microscope. (b) High-resolution TEM images of Ag nanoparticles recorded during an in situ heating experiment while increasing the temperature from 25°C to 200°C with dynamic control activated. The same field of view is preserved throughout the temperature ramp. (c) Corresponding images acquired without specimen drift compensation, illustrating the large specimen displacement induced by the temperature ramp. The red dotted rectangle corresponds to the initial field of view.*

### a. Example on an *in situ* heating HR-TEM experiment

A first experiment was carried out in TEM mode to evaluate the performance of the dynamic control during an *in situ* heating experiment, a particularly challenging situation owing to the large specimen drifts induced by thermal stimulus. The experiment was performed on a double aberration-corrected JEM-ARM200F microscope equipped with a OneView camera. Silver nanoparticles deposited on a MEMS-based heating holder (Fusion Holder, Protochips Inc.) to locally apply a temperature ramp. High-resolution TEM observations were performed while increasing the temperature from 25°C to 200°C at a rate of 0.5°C.$s^{-1}$ with and without dynamic control. During the temperature ramp, frames were acquired every 0.5 s. The full frame series can be found in Supplementary Data.

Figures 4b and 4c present images acquired at 25°C, 125°C and 200°C with and without specimen drift compensation, respectively. When specimen drift compensation is activated (Figure 4b), the same 100 × 100 $nm^2$ field of view is preserved throughout the entire temperature ramp without any user intervention other than manual focus compensation. The feedback loop operates sufficiently fast and with sufficient positioning precision to preserve atomic-resolution information from one frame to the next. Real-time observation and recording of crystallographic reorientations and moiré contrast variations induced by temperature are therefore possible. In contrast, without specimen drift compensation (Figure 4c), the total specimen displacement reaches approximately 40 nm (1700 pixels), reducing the initial field of view by nearly 50 % and preventing atomic-resolution observations from being maintained throughout the experiment. Manual specimen compensation would be both less precise and inevitably would degrade the effective spatial resolution of the acquisition.

This experiment highlights the relevance of the dynamic control for *in situ* electron microscopy by demonstrating its ability to actively stabilize the specimen position at the atomic scale despite large thermally induced drifts and without compromising spatial resolution. The ultimate limitation of the approach lies in the drift dynamics themselves, which must remain compatible with both the acquisition time of individual frame and the response time of the feedback loop.

### b. Quantitative high-resolution STEM experiments

The second experiment was performed in STEM mode on a different JEM-ARM200F microscope in order to evaluate the performance of our approach for high-resolution acquisitions using HAADF and ABF detectors coupled to a fast-scan system (DigiScan3, Gatan Inc.). No intentional specimen drift was introduced during the experiment. The investigated specimen consists of a $La_{2/3}Sr_{1/3}MnO_3$ (LSMO)/$BaTiO_3$ (BTO)/LSMO trilayer deposited on a $SrTiO_3$ (STO) substrate and observed in cross-sectional geometry.

A first experiment was carried out at the STO/LSMO interface over a 17 × 17 $nm^2$ field of view. A total of 270 HAADF frames were acquired over 540 s (9 min) at a rate of one frame every 2 s (Figure 5a). Each frame consists of 1024 × 1024 pixels acquired with a dwell time of 0.8 μs. The horizontal $x$ and vertical $y$ image axes correspond to the fast- and slow-scan directions, respectively, as indicated in Figure 5a. Prior to the acquisition sequence, the specimen drift compensation was activated using an atomic unit cell as the reference feature for drift measurements. The feedback unit remained active during the acquisition of the first 90 frames before being deactivated for the following 90 frames and subsequently reactivated for the last 90 frames.

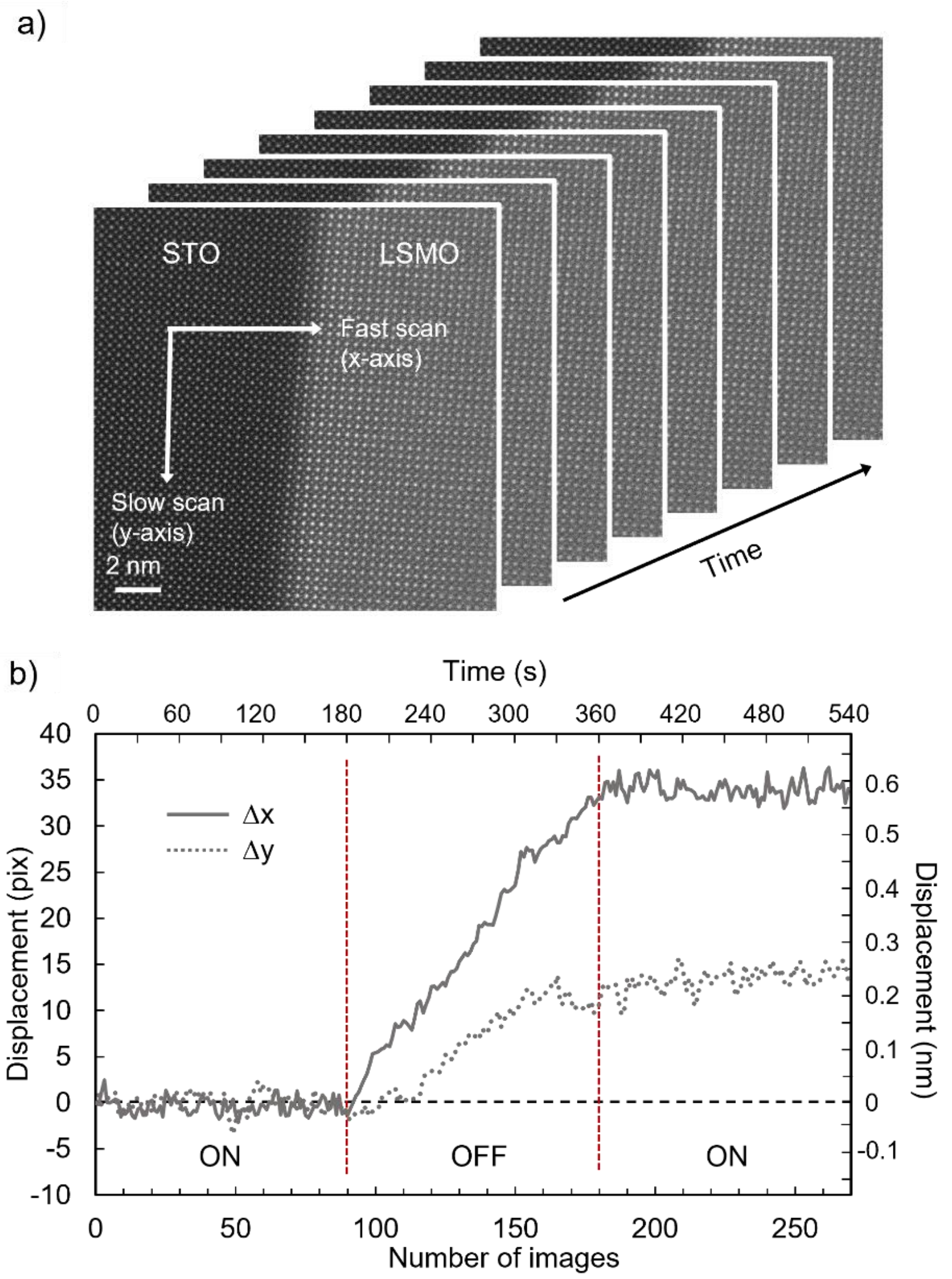

***Figure 5.*** *(a) Acquisition sequence of 270 high-resolution HAADF-STEM frames (1024 × 1024 pixels) acquired over 540 s at the STO/LSMO interface. The specimen drift compensation was activated during the first and last thirds of the experiment and deactivated during the intermediate third. (b) Specimen drift measured by post-processing software, demonstrating the effectiveness of the feedback unit in maintaining the specimen position during long STEM acquisitions.*

Figure 5b presents the specimen drift $\Delta x$ and $\Delta y$ measured in post-processing using a dedicated software for numerical drift correction of image series. The first frame has been used as reference frame to evaluate the drift (see Supplementary Data). When the feedback unit is active, only a residual drift is observed, with a mean value about 0.4 pixel (7 pm) for the first 90 images and a standard deviation of approximately 1 pixel (17 pm). This residual displacement originates from both the accuracy of the feedback loop and the measurement uncertainty associated to the post-processing image registration software. When the feedback unit is deactivated, a nearly linear specimen drift is observed, predominantly along the $x$ image axis, reaching approximately 35 pixels (0.6) nm before the feedback unit is reactivated. Although relatively small, this displacement is more than sufficient to limit the maximum acquisition time compatible with atomic-resolution STEM imaging under conventional operating conditions. When the feedback unit is reactivated, the specimen position is stabilized once again with the same residual displacement.

To quantitatively compare stage-based specimen stabilization with conventional post-acquisition image correction, the first 90 frames acquired with the feedback unit on were directly summed (Figure 6a). The 90 frames acquired without live compensation were aligned numerically using the drift values measured in Figure 5b before being summed to produce Figure 6b. Finally, Figure 6c corresponds to a conventional single STEM acquisition performed over 40 s using a dwell time of 38 µs. Enlarged views of the interface are presented for each image. Figures 6a and 6b, displayed using identical contrast levels, exhibit very similar image quality. The atomic columns are clearly resolved and the contrast differences between the various cationic sites are preserved. Scan distortions, averaged out over a large number of rapidly acquired frames, are absent in both images. In contrast, they remain clearly visible in the conventional acquisition (Figure 6c), where they give rise to the characteristic "hairy atom" artefacts. The most apparent difference between Figures 6a and 6b is the reduction of the effective field of view associated with the post-processing alignment, as indicated by the dashed red rectangle in Figure 6b.

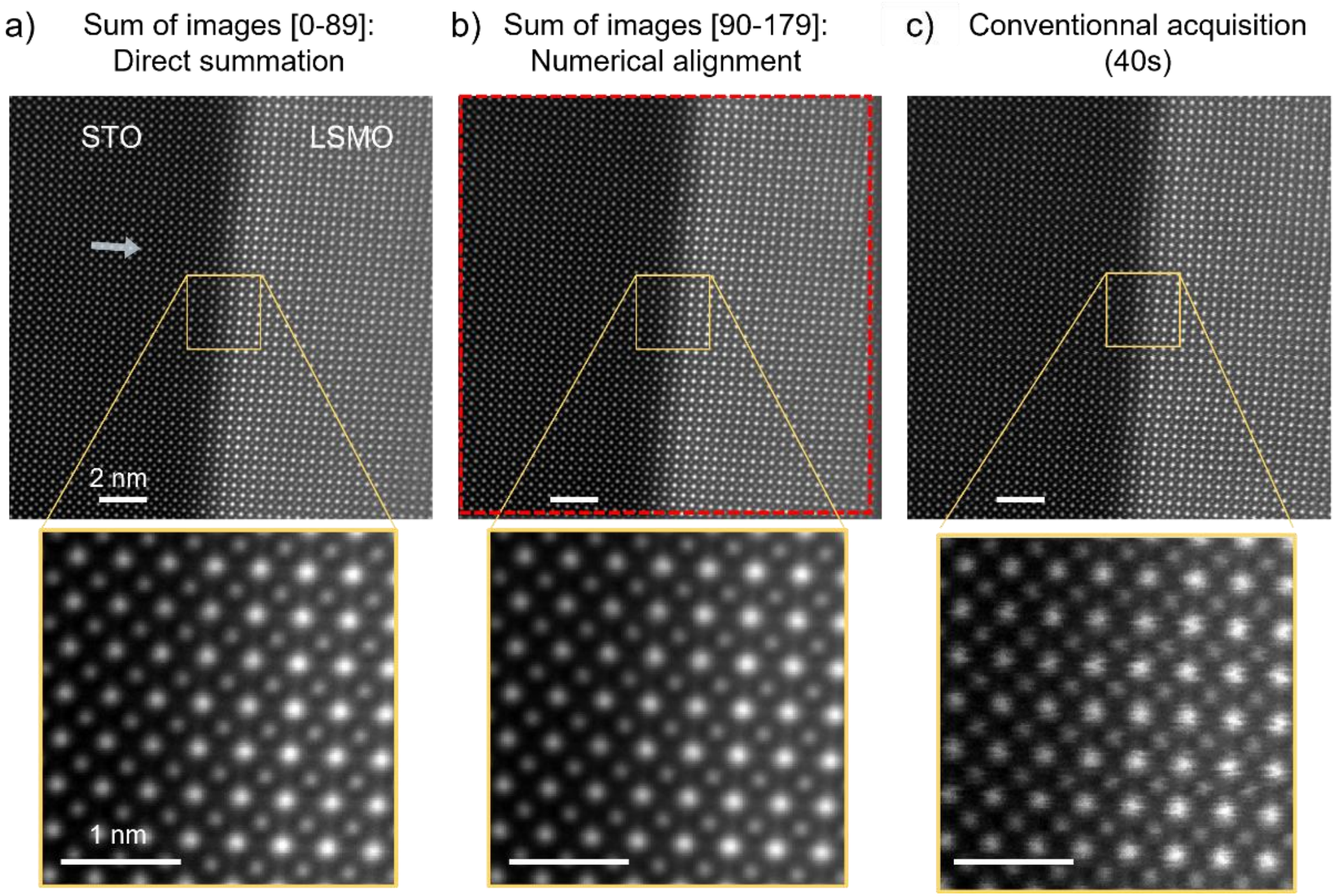

***Figure 6.*** *(a) HAADF-STEM image obtained by real-time summation of 90 frames acquired while the dynamic control was active. (b) Image obtained by post-processing registration and summation of 90 frames acquired without specimen drift compensation. The dashed red rectangle indicates the effective reduction of the field of view resulting from the numerical alignment. (c) Conventional HAADF-STEM acquisition performed using a single 40 s scan. Enlarged views of the STO/LSMO interface are shown for each image.*

Stage-based specimen stabilization implemented through the dynamic control therefore produces an image quality at least equivalent to that obtained using post-processing registration of frame series. A more quantitative comparison is provided in Figure 7a, which presents intensity profiles extracted from Figures 6a and 6b along the direction indicated by the white arrow in Figure 6a. Both profiles are similar. However, the image obtained after numerical alignment may exhibit slightly reduced peak amplitudes together with a subtle broadening of the atomic peaks. This effect originates from the sub-pixel image interpolation required during the numerical registration procedure, which effectively acts as a low-pass digital filter on the image intensity distribution as illustrated in Figure 7b. The image extracted from the frame series acquired without live compensation is shifted numerically by a non-integer number of pixels according to the measured drift vector. The resulting image clearly exhibits an intensity smoothing effect. Indeed, the pixels are no longer independent which distorts the counting statistics. To avoid this, it is necessary to implement advanced interpolation methods.

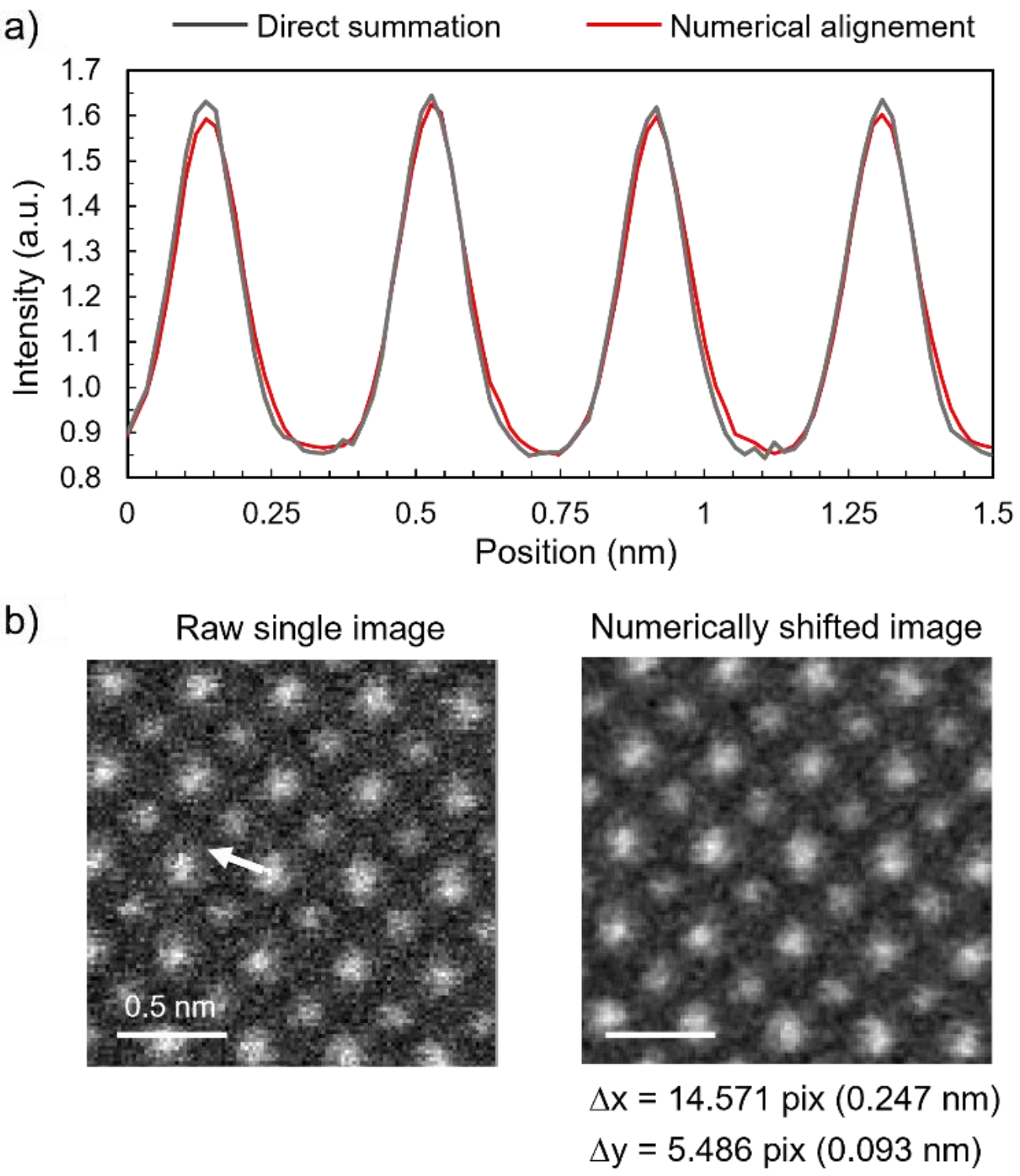

***Figure 7.*** *(a) Intensity profiles extracted from Figures 6a and 6b along the direction indicated by the white arrow in Figure 6a. (b) Illustration of the intensity smoothing induced by sub-pixel interpolation when a numerical drift is applied with non-integer pixel displacement. The right image corresponds to the left image numerically displaced along the white arrow with the specified values.*

These results demonstrate that stage-based specimen stabilization avoids the introduction of interpolation artefacts, in addition to reducing both the computational cost and the amount of data generated during long acquisitions. However, the practical limit of the feedback loop is strongly influenced by the frame acquisition time, in particular in high-resolution STEM. The loop can compensate specimen displacement between successive frames, but it cannot recover spatial information already degraded by drift during the acquisition of an individual frame. Under the present DigiScan3 conditions, one correction every 2 seconds can be achieved. Specimen drifts of a few nanometres per minute can therefore be compensated while preserving atomic-resolution imaging. Faster drift would require shorter frame times, for example by reducing the number of scanned pixels, at the expense of spatial sampling and/or field of view.

When frame summation is performed in parallel with the feedback-unit operation, the dynamic control directly produces a single image suitable for the analysis of structural distortions or changes in crystallographic orientation. An example is provided in Figure 8, which presents a STEM-HAADF observation of a $PbZr_{0.2}Ti_{0.8}O_3$ (PZT) layer. The investigated region consists of a *c* domain surrounded by two *a* domain.

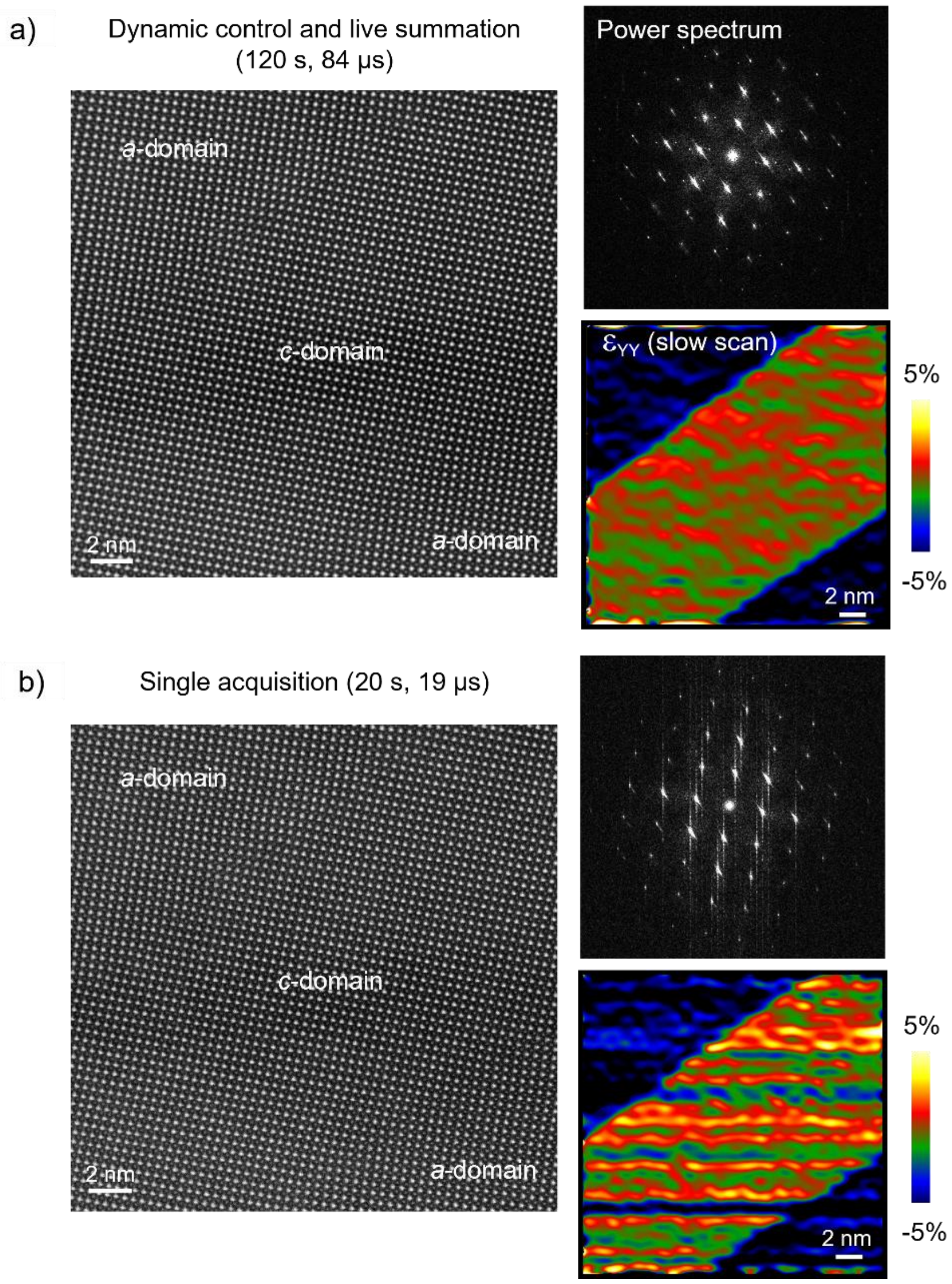

***Figure 8.*** *HAADF-STEM images of a PbZr0.2Ti0.8O3 layer acquired (a) using the dynamic control (120 s for a total pixel time of 84 µs) and (b) by conventional STEM acquisition (20 s with a pixel time of 19 µs). Corresponding Fourier transforms and geometrical Phase Analysis (GPA) strain maps calculated along the slow-scan direction for each HAADF image highlight the suppression of scan distortions when the dynamic control is used by real-time frame summation and specimen stabilization.*

Two images were acquired from this same region. The first was obtained by summing in real time 60 frames delivered every 2 s by the HAADF detector (see Supplementary Data) while the feedback unit was active (Figure 8a). Considering a dwell time of 1.4 µs per frame, the total pixel time is 84 µs. The second was obtained using a single conventional acquisition with an acquisition time of 20 s and a total pixel time of 19 µs (Figure 8b). As shown by the Fourier transforms of the corresponding HAADF images, scan distortions are absent from the image acquired using the dynamic control, whereas they are clearly visible in the conventionally acquired image. These distortions generate substantial noise in the strain maps, here calculated by Geometric Phase Analysis (GPA)[42,43] along the slow-scan direction. Their suppression, combined with a higher electron dose, results in strain maps with a substantially improved signal-to-noise ratio, without requiring any additional numerical correction. Furthermore, the rapid scanning used during the dynamic control limits specimen contamination compared with a conventional slow scan, thereby allowing longer total exposure times.

The final example illustrates the coupling of the specimen drift compensation with a STEM acquisition involving multiple detectors. As discussed above, drift measurements are more accurate when the image feature used by the sensor exhibits strong contrast. HAADF images generally provide sufficient contrast for reliable drift compensation, even at short frame times. This is not always the case for signals recorded by other detectors, particularly when their signal-to-noise ratio becomes too low at short frame times. This limitation may arise, for example, with an ABF detector used to visualize light elements. Dynamic control nevertheless allows images to be acquired with such detectors by measuring the specimen drift from a higher-contrast or less noisy image simultaneously recorded by another detector.

Figure 9 presents two images acquired simultaneously using HAADF and ABF detectors while the dynamic control was used. The specimen was the LSMO/BTO/LSMO trilayer described above. The total acquisition time of 150 s corresponds to the real-time summation of 75 frames acquired every 2 s by the acquisition unit operating in parallel with the drift-compensation feedback unit. At the end of the acquisition, only two final images, one for each detector, were stored. The specimen drift was measured exclusively from the HAADF data stream. Three regions were selected and enlarged to illustrate the effectiveness of this simultaneous acquisition scheme. The HAADF images exhibit high image quality, with the different cationic sites clearly identified from their intensity differences. The ABF images show a similarly high level of quality and clearly reveal the oxygen sublattice surrounding the cations. Such information is particularly valuable for measurements of atomic polarization in ferroelectric materials. The dislocation structure in region C is particularly detailed and clearly resolved in the ABF image.

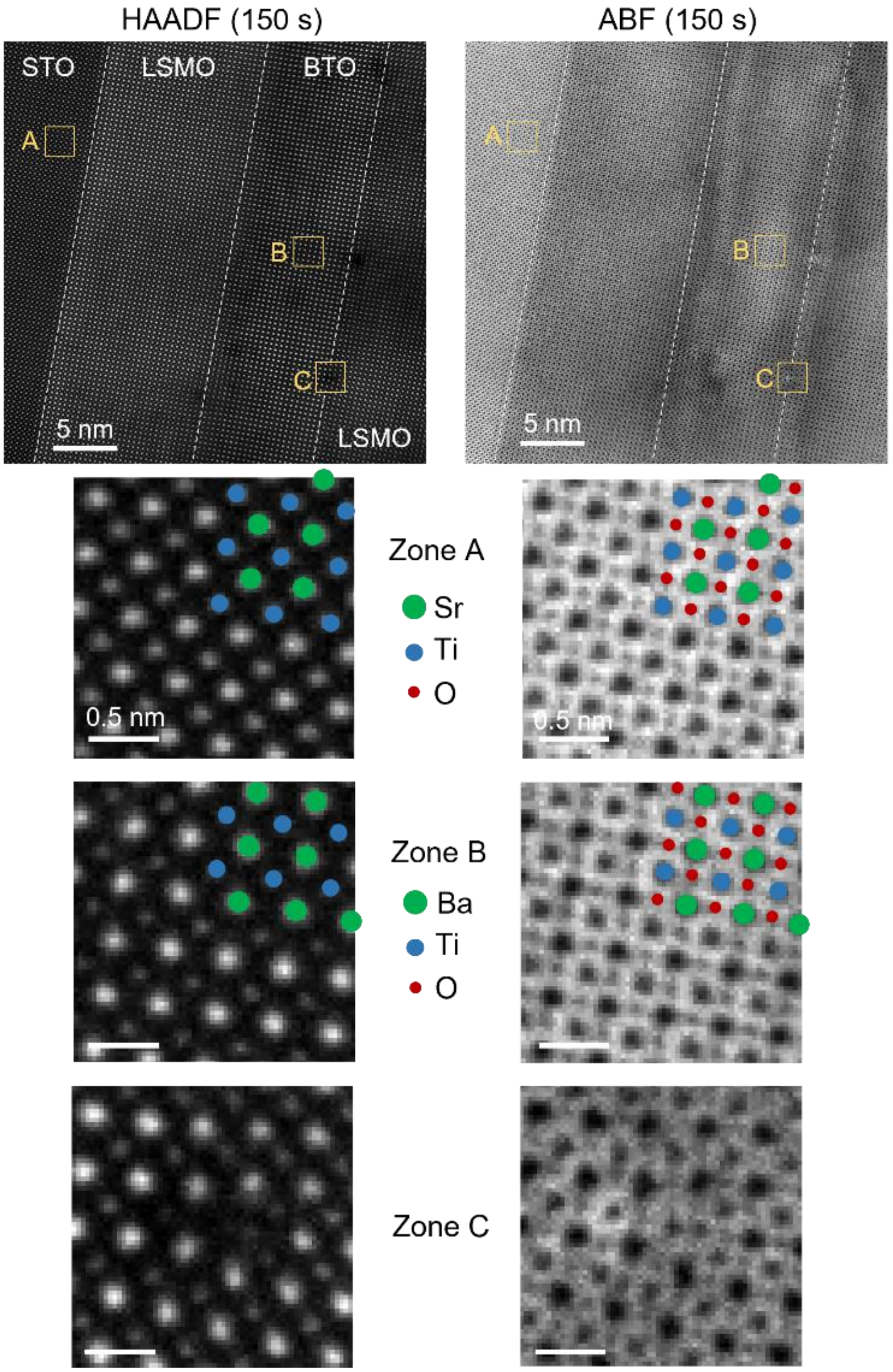

***Figure 9.*** *Simultaneously acquired HAADF and ABF STEM images of an LSMO/BTO/LSMO trilayer obtained by real-time summation of 75 frames acquired over a total acquisition time of 150 s using the dynamic control. Specimen drift was measured exclusively from the HAADF data stream. Enlarged views of selected regions illustrate the high image quality of the atomic contrast in both detectors.*

The combination of the dynamic control with multi-detector acquisition is not restricted to HAADF and ABF imaging. The same approach can readily be extended to a wide range of electron microscopy techniques, including electron energy-loss spectroscopy (EELS) and energy-dispersive X-ray spectroscopy (EDS) for chemical mapping, integrated differential phase contrast (iDPC) imaging, differential phase contrast (DPC) and four-dimensional STEM (4D-STEM) using segmented or pixelated detectors, as well as ptychography. More generally, the dynamic control is compatible with any detector or combination of detectors capable of simultaneously providing the information required for specimen drift measurement and the experimental signal of interest.

## Conclusion

Continuous improvements in electron microscopy instrumentation have considerably increased the spatial resolution, sensitivity and quantitative capabilities of modern TEM and STEM experiments. These advances have simultaneously made instrumental instabilities, and particularly specimen drift, one of the main factors limiting long-exposure observations and quantitative measurements. In this work, we have shown that specimen drift can be compensated through feedback control of the stage to improve the experimental data acquired at the microscope. The framework has been successfully implemented on several microscopes from different manufacturers and demonstrated under a wide range of experimental conditions, from medium-resolution to atomic-resolution and TEM to STEM. Depending on the positioning capabilities of the specimen stage, stabilization ranging from the nanometer down to the picometer scale has been achieved, enabling virtually unlimited atomic-resolution observations by continuously maintaining the specimen within the same field of view.

The specimen drift compensation directly produces high-quality images without acquisition of large datasets, artefacts from numerical alignment, loss of field of view or additional computational cost. Increasing the effective exposure time improves the signal-to-noise ratio in both TEM and STEM, while the rapid acquisition of individual frames suppresses random scan distortions and significantly improves the quality of quantitative analyses. The framework is compatible with simultaneous multi-detector acquisitions and with long-duration *in situ* experiments.

Nevertheless, every experiment has its own set of unique challenges and drift compensation is not guaranteed to work every time. The greatest limitation is, of course, the speed of the drift but the critical amount depends on the experimental constraints. For example, a greater drift speed can be accommodated during medium-resolution imaging (holography or conventional TEM) than during high-resolution imaging of the atomic lattice. Similarly, the higher frame rate of TEM imaging with respect to STEM, raises the critical drift rate. As a general rule, the total drift during a single frame should be lower than the desired spatial resolution.

Neither can the frame rate be increased indefinitely to compensate the drift speed. Each frame will have a lower dose and hence the accuracy of the sensor measuring the drift will surpass the required resolution at some point. A compromise must be made. Similarly, reducing the pixel size of the scan for STEM imaging will eventually defeat the purpose of the experiment by reducing the field of view too adversely. And ultimately, once the frame rate is no longer the limiting factor, the response time of the microscope will take the relay. This is not just a question of the mechanical response of a stage to a solicitation but the electronics and on-board computer of the particular microscope. We can conceive that a modern microscope responds more quickly than an older instrument and that a piezo stage reacts more quickly than a purely mechanical mechanism. It is difficult to give definite numbers. We anticipate that feedback control through beam tilts and image shifts are faster than mechanical stage movements, though current stabilisation in the electromagnetic lenses needs to be taken into account. Finally, in our opinion, the calculation time for the corrections will never be the limiting factor given current computer power.

Implemented as a DigitalMicrograph plugin, the dynamic control requires no modification of the microscope hardware and can be readily deployed on modern electron microscopes providing access to detector data streams and microscope control interfaces. Beyond specimen drift compensation, the same principles can be extended to other controllable experimental parameters, making the dynamic control a flexible framework for the real-time stabilization of electron microscopy experiments.

## Acknowledgements

The authors acknowledge funding of the French National Research Agency for the POLARYS project (ANR-23-CE42-0011) and the "Investissement d'Avenir" program reference Nos. ANR-10-EQPX-38-01 and 11-IDEX-0002, the "Conseil Regional Midi-Pyrénées" and the European FEDER-FSE for financial support within the CPER program. The authors also acknowledge funding from the European Union under grant agreement no. 101094299 (IMPRESS). The authors would also like to thank Cécile Genevois (CNRS, CEMHTI, Univ, Orléans, France), Corine Bouillet (Plateforme MACLE-CVL, Orléans, France) and Martien Den Hertog (Univ. Grenoble Alpes, CNRS, Grenoble INP, Institut Néel, Grenoble, France) for providing access to JEOL microscopes, and J.B. Moussy, A. Barbier (LNO-SPEC, CEA Saclay, France), and S. Matzen (Univ. Paris-Saclay, CNRS, C2N, Palaiseau, France) for the test sample used in this work.

## Appendix A. Supplementary Data

Raw frames from Figures 5 and 8 extracted from the detector frame rate and the corresponding final images obtained from direct summation and specimen compensation thanks to the dynamic control. Comparison with conventional single acquisition.

Two videos corresponding to the series of 350 frames recorded during the *in situ* heating experiment with and without dynamic control (Figure 4b). Speed × 10.